\documentclass[conference]{IEEEtran}
\usepackage{cite}
\usepackage{amsmath,amssymb,amsfonts}
\usepackage{algorithmic}
\usepackage{graphicx}
\usepackage{textcomp}
\usepackage{xcolor}
\usepackage[hyphens]{url}
\usepackage{multirow}
\def\BibTeX{{\rm B\kern-.05em{\sc i\kern-.025em b}\kern-.08em
    T\kern-.1667em\lower.7ex\hbox{E}\kern-.125emX}}
\usepackage[braket, qm]{qcircuit}
\usepackage{subcaption}

\title{QASMTrans: A QASM based Quantum Transpiler Framework for NISQ Devices}

\author{
\IEEEauthorblockN{
    Fei Hua\IEEEauthorrefmark{1}\IEEEauthorrefmark{2}, 
    Meng Wang\IEEEauthorrefmark{1}\IEEEauthorrefmark{5}, 
    Gushu Li\IEEEauthorrefmark{3},
    Bo Peng\IEEEauthorrefmark{1},
    Chenxu Liu\IEEEauthorrefmark{1},
    Muqing Zheng\IEEEauthorrefmark{1},
    \\
    Samuel Stein\IEEEauthorrefmark{1},
    Yufei Ding\IEEEauthorrefmark{4},
    Eddy Z. Zhang\IEEEauthorrefmark{2},
    Travis S.~Humble\IEEEauthorrefmark{6}, 
    Ang Li\IEEEauthorrefmark{1}
}
\IEEEauthorblockA{
    \IEEEauthorrefmark{1}\textit{Pacific Northwest National Laboratory}, 
    \IEEEauthorrefmark{2}\textit{Rutgers University},\\
    \IEEEauthorrefmark{3}\textit{University of Pennsylvania}, 
    \IEEEauthorrefmark{4}\textit{University of California San Diego},\\
    \IEEEauthorrefmark{5}\textit{The University of British Columbia},
    \IEEEauthorrefmark{6}\textit{Oak Ridge National Laboratory}
}}

\begin{document}
\maketitle
\thispagestyle{plain}
\pagestyle{plain}


 \begin{abstract}

The success of a quantum algorithm hinges on the ability to orchestrate a successful application induction. Detrimental overheads in mapping general quantum circuits to physically implementable routines can be the deciding factor between a successful and erroneous circuit induction. In QASMTrans, we focus on the problem of rapid circuit transpilation. Transpilation plays a crucial role in converting high-level, machine-agnostic circuits into machine-specific circuits constrained by physical topology and supported gate sets. The efficiency of transpilation continues to be a substantial bottleneck, especially when dealing with larger circuits requiring high degrees of inter-qubit interaction. QASMTrans is a high-performance C++ quantum transpiler framework that demonstrates 10-369$\times$ speedups compared to the commonly used Qiskit transpiler. We observe speedups on large dense circuits such as `uccsd\_n24' and `qft\_n320' which require  
$\mathcal{O}(10^6)$ gates. QASMTrans successfully transpiles the aforementioned circuits in {69s} and {31s}, whilst Qiskit exceeded an hour of transpilation time. With QASMTrans providing transpiled circuits in a fraction of the time of prior transpilers, potential design space exploration, and heuristic-based transpiler design becomes substantially more tractable. QASMTrans is released at http://github.com/pnnl/qasmtrans.

 \end{abstract}
\section{Introduction}

The past decade has witnessed tremendous development in \emph{Noisy Intermediate-Scale Quantum} (NISQ) computers \cite{clarke2008superconducting, rigetti2012superconducting}, where a few hundred physical qubits are available with relatively limited coherence times and high error rates. These NISQ machines, while offering great potential, are constrained by various factors such as non-trivial noise~\cite{maciejewski2020mitigation}\cite{tannu+:micro19}, limited connectivity~\cite{chamberland2020topological} and machine-specific basis gate sets~\cite{lin2022let}. Due to the limited qubit number and short coherence time, effectively mapping application circuits to the constrained NISQ machine poses a considerable challenge and can significantly impact the fidelity of the execution results.

Transpilation is the specific terminology referring to the compilation process of transforming a high-level quantum circuit into an equivalent circuit that is compatible with the specifications of a quantum device, including: basis gate set, topology of the quantum chip, timing constraints, fidelity of operations, etc. The goal of a transpiler is to perform this transformation while minimizing the impact on the functionality of the circuit and optimizing its performance delivery.

Several attempts on quantum transpilation have already been made by the community (see a summary in Section~\ref{sec:related_work}), but there are still technical gaps. On the one hand, commercial transpilers such as those embedded in Qiskit~\cite{IBMQiskit} and Cirq~\cite{cirq} provide comprehensive functionalities, but are typically slow, especially for those deep circuits arising in practical quantum applications such as chemistry~\cite{kauzmann+:qcbook13, cao+:cr19}, optimization~\cite{dunjko+:rpp18, wichert2020principles} and nuclear physics~\cite{stetcu2022projection, holmes2022quantum}. Additionally, the slow transpilation speed limits their capability to explore larger design space and integrate more advanced but expensive optimizations. This is especially the case when dynamic circuit generation and transpilation is needed, such as in variational quantum algorithms (VQAs)~\cite{cerezo2021variational, stein2022eqc} and when optimized to mitigate state-dependent bias at runtime \cite{tannu2019mitigating}.

On the other hand, most of the research studies in academia have focused on specific transpilation techniques, such as gate decomposition, circuit optimization, mapping and routing, etc.~\cite{li+:asplos19}\cite{zulehner+:date18:}\cite{zhang+:asplos21}. These approaches lack end-to-end demonstrations and are often implemented and validated by embedding into or replacing part of the Python-based commercial frameworks such as Qiskit and Cirq. Consequently, they are also constrained by the limitations of the underlying frameworks, such as slow speed, difficulties in launching large circuits, binding to certain device features, lack of flexibility, and frequent interface upgrades, etc. 

\begin{figure}[htb]

  \centering
  \includegraphics[width=0.95\linewidth]{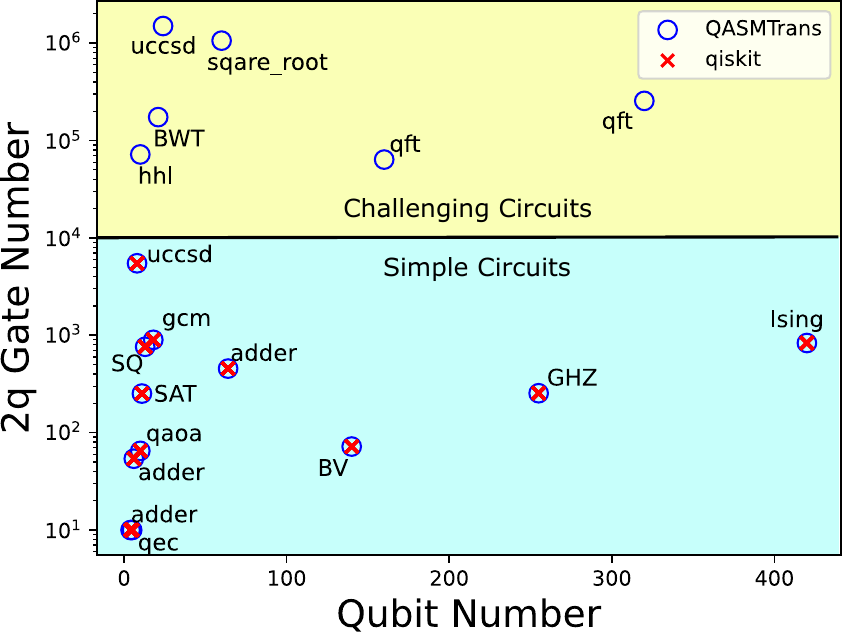}
  \caption{QASMTrans is designed to transpile challenging deep circuits from QASMBench~\cite{li2023qasmbench}, while broadly used tools such as Qiskit cannot finish within an hour.}
  \label{fig:compilation_time}
\end{figure}

In this paper, we present QASMTrans, an end-to-end, self-contained, light-weight quantum transpiler entirely realized in C++ for effectively parsing and compiling large QASM circuits. QASMTrans comprises four major components: 


\begin{enumerate}
    \item An {\textbf{\emph{IO}}} module that uses a QASM Paser for parsing an input OpenQASM file, and translating it into a structure acting as the internal intermediate representation (IR). The output will export the transpiled QASM circuits for a particular NISQ device, such as those provided by IBMQ, Rigetti, IonQ, Quantinuum, etc.
    \item A {\textbf{\emph{Configuration}}} module for preparing the coupling graph of the device, generating the DAG for the circuit, and decomposing the 3-qubit gates into 1-qubit and 2-qubit gates.
    \item An {\textbf{\emph{Optimization}}} module for the various optimization passes. This includes decomposition into basis gates, routing, and mapping. These passes are made with respect to the topology, basis gate set, fidelity, and features of the circuit. The goal of the backend optimization is to allow the circuits to run more efficiently on the targeted NISQ devices or simulators.
    \item A main \textbf{\emph{Transpiler}} component to do the routing and mapping and also decompose into basis gates based on specific NISQ devices.
\end{enumerate}

QASMTrans is primarily designed as an open-source transpiler infrastructure serving as a baseline for implementing and validating advanced transpilation technologies while supporting novel devices and computation models. We evaluate QASMTrans using diverging circuits with some of them being quite challenging (from 4 to 420 qubits, and from 10 to 2.2M gates, see Figure~\ref{fig:compilation_time}) from QASMBench~\cite{li2023qasmbench}. Remarkably, most of the benchmarks can be completed within a few seconds. Even the largest and most demanding benchmark that Qiskit cannot finish within an hour, can be transpiled by QASMTrans in 69s. This work thus makes the following main contributions:

\begin{itemize}
    \item We propose an end-to-end, self-contained, light-weighted opensource quantum compiler in C++ that can significantly reduce the transpilation time for a wide range of applications, improving the efficiency of quantum computations on NISQ devices.
    \item QASMTrans is equipped with optimization techniques for generating specific basis gates towards different target machines or classical simulators. 
    \item Through comprehensive experiments and analysis over multiple quantum platforms, we show that QASMTrans can transpile circuits with comparable fidelity on real NISQ devices from Rigetti, IBMQ, IonQ and Quantinuum, but at a much faster speed compared to existing transpilers such as Qiskit and MQT-Qmap \cite{wille2023mqt}.  
\end{itemize}

The remainder of this paper is structured as follows: Section~\ref{sec:background} provides background information. Section~\ref{sec:transpiler} presents the QASMTrans transpiler. Section~\ref{sec:evaluation} shows the evaluation results. Section~\ref{sec:related_work} summarizes related work about quantum transpilation. Section~\ref{sec:conclude} concludes.

\section{background} \label{sec:background}
\begin{table*}[!t]
\centering\scriptsize
\caption{OpenQASM gate definition (5 basic gates + 11 standard gates + 18 composition gates).}
\begin{tabular}{|c|l|c|l|c|l|}
\hline
\textbf{Gates} & \textbf{Meaning} & \textbf{Gates} & \textbf{Meaning} & \textbf{Gates} & \textbf{Meaning}  \\ \hline
\texttt{U3} & 3 parameter 2 pulse 1-qubit & \texttt{TDG} & conjugate of sqrt(S) &  \texttt{CRZ} & Controlled RZ rotation  \\ \hline
\texttt{U2} & 2 parameter 1 pulse 1-qubit  & \texttt{RX} & X-axis rotation    &  \texttt{CU1} & Controlled phase rotation   \\ \hline
\texttt{U1} & 1 parameter 0 pulse 1-qubit  & \texttt{RY} & Y-axis rotation  &  \texttt{CU3} & Controlled U3   \\ \hline
\texttt{CX} & Controlled-NOT & \texttt{RZ} & Z-axis rotation     &  \texttt{RXX} & 2-qubit XX rotation   \\ \hline
\texttt{ID} & Idle gate or identity & \texttt{CZ} & Controlled phase    & \texttt{RZZ} & 2-qubit ZZ rotation   \\ \hline
\texttt{X} & Pauli-X bit flip   & \texttt{CY} & Controlled Y   & \texttt{RCCX} & Relative-phase CXX    \\ \hline
\texttt{Y} & Pauli-Y bit and phase flip   & \texttt{SWAP} & Swap   & \texttt{RC3X} & Relative-phase 3-controlled X    \\ \hline
\texttt{Z} & Pauli-Z phase flip   &  \texttt{CH} & Controlled H   & \texttt{C3X} & 3-controlled X  \\ \hline
\texttt{H} & Hadamard   &  \texttt{CCX} & Toffoli  & \texttt{C3XSQRTX} & 3-controlled sqrt(X)   \\ \hline
\texttt{S} & sqrt(Z) phase  &  \texttt{CSWAP} & Fredkin   & \texttt{C4X} & 4-controlled X  \\ \hline
\texttt{SDG} & conjugate of sqrt(Z)  & \texttt{CRX} & Controlled RX rotation & &   \\ \hline
\texttt{T} & sqrt(S) phase   & \texttt{CRY} & Controlled RY rotation & & \\ \hline
\end{tabular}
\label{tab:gates}
\end{table*}
\subsection{Noisy Intermediate-Scale Quantum (NISQ)}

NISQ systems refer to near-term quantum platforms featuring fifty to less than a thousand physical qubits \cite{preskill2018quantum}. These qubits are fabricated based on various technologies, such as superconducting \cite{clarke2008superconducting, rigetti2012superconducting}, trapped-ion \cite{cirac1995quantum, leibfried2003quantum}, photonic \cite{o2009photonic, aspuru2012photonic}, spin qubits \cite{pla2012single, maurand2016cmos}, neutral atoms \cite{briegel2000quantum, henriet2020quantum}, etc. To accomplish the execution of a circuit, the physical qubits need to stay coherent for a sufficiently long time. However, before all the circuits can be executed on the real quantum machine, it must 1) fit the basis gates of the quantum machine and 2) meet the coupling constraints of the machine topology.

\subsubsection{Basis Gates}

Each NISQ device has its own basis gate set \cite{ibm, ionq, rigetti, quantinuum}, known as the \emph{quantum instruction set architecture} (QISA). It defines the basic operations that are physically supported by the underlying platform. During quantum transpilation, all the logic gates will be decomposed and transpiled into gate sequences purely formed by basis gates.  Table~\ref{tab:basis_gates} shows the basis gate set for IBMQ, Rigetti, IonQ and Quantinuum devices. Typically, quantum device vendors only provide profiling or calibration data for the basis gates (per qubit or system-wide average), including T1, T2, duration, fidelity, etc. These basis gates also represent the operations to be implemented by a classical simulator.

\subsubsection{Topology}

\begin{figure}[!t]
\minipage{0.33\columnwidth}
\includegraphics[width=1.0\columnwidth]{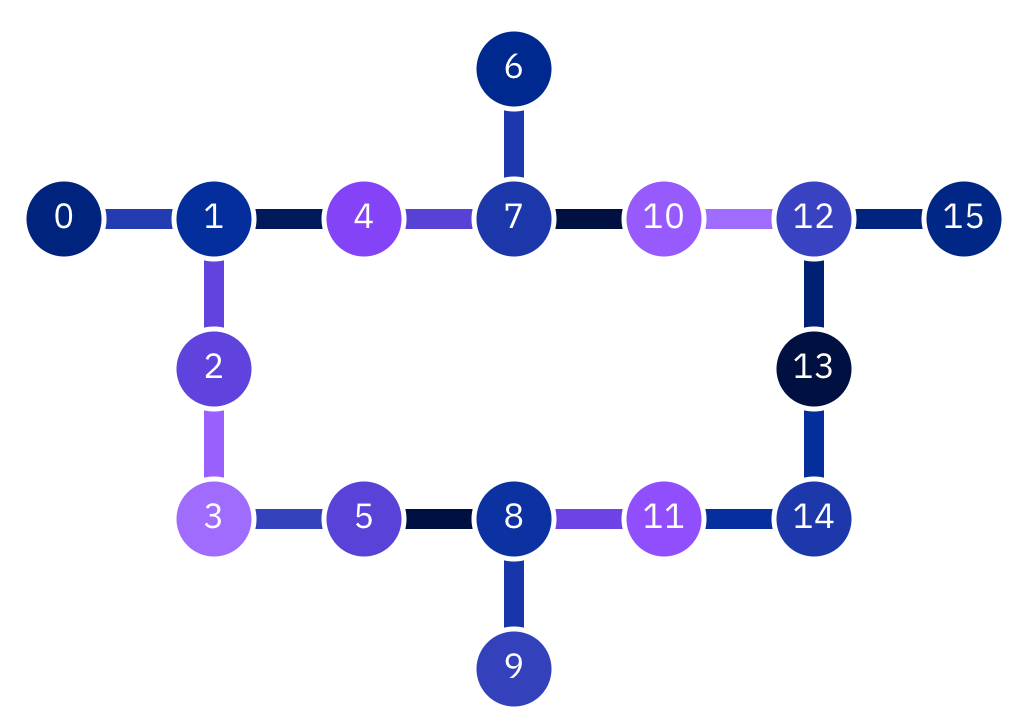}
\endminipage\hfill
\minipage{0.37\columnwidth}
\includegraphics[width=1.0\columnwidth]{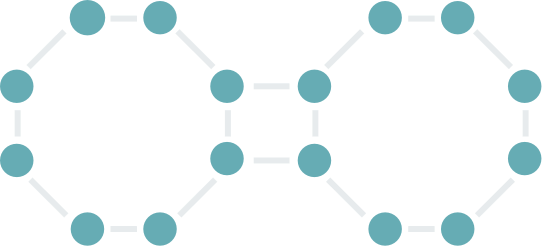}
\endminipage\hfill
\minipage{0.25\columnwidth}
\includegraphics[width=1.0\columnwidth]{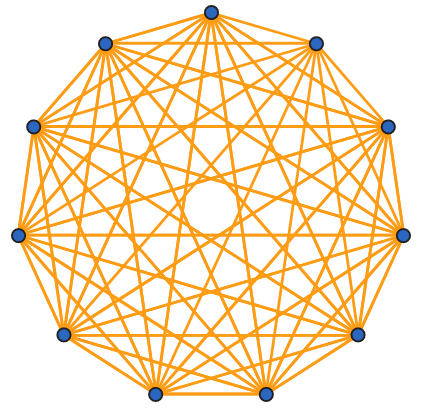} 
\endminipage
\caption{NISQ device topology: IBMQ-Guadalupe (left), Rigetti-Aspen (middle) and IonQ-QPU (right).}
\label{fig:ibmq_topo}
\end{figure}

Physical qubits in a quantum processor are interconnected. In a quantum device, the 1-qubit gates are directly performed on individual qubits. The 2-qubit gates, however, have to be performed on a qubit-pair that is interconnected. This is especially the case for superconducting devices (e.g., IBMQ and Rigetti), where the connectivity of qubits follow a certain topology, as shown in Figure~\ref{fig:ibmq_topo}. The topology thus limits the sites where two-qubit gates can be performed: if a two-qubit gate is desired for remote qubits, a series of \texttt{SWAP} gates are required to physically move the two qubits to a connected tuple following the path defined by the topology, known as \emph{routing}. \texttt{SWAP} gates are costly, usually achieved through three \texttt{CNOT} or \texttt{CX} gates. 

These extra \texttt{SWAP}s are one of the major factors contributing to deep circuits and considerable noise for superconducting devices, as compared to contemporary small-scale trapped-ion devices practicing all-to-all connectivity (see Figure~\ref{fig:ibmq_topo}). Our previous study \cite{stein2022quclassi} shows that, for a 17-gate variational circuit, from the 5-qubit IBMQ Cairo to the 5-qubit IonQ QPU, a fidelity increase from 72\% to 80\% (ideally 97.8\%) has been observed. This is mainly due to the 7 extra \texttt{SWAP} gates when transpiling to comply with the topology of IBMQ Cairo.

\subsection{QASM}

OpenQASM (Open Quantum Assembly Language, we particularly refer to OpenQASM~2.0 in this work) \cite{cross2017open}, also known colloquially as QASM, is an intermediate representation (IR) of quantum instructions. QASM acts as a unified low-level assembly language for IBMQ and other quantum machines. Many of these NISQ devices, accessible through the IBMQ network \cite{ibm}, have been widely explored by existing works. 

Table~\ref{tab:gates} lists the types of gates that are defined in the QASM specification (i.e., the "\texttt{qelib1.inc}" header file) \cite{cross2017open}. Within these gates, the first five, i.e., \texttt{U3}, \texttt{U2}, \texttt{U1}, \texttt{CX}, and \texttt{ID}, are \emph{basic gates} that are expected to be supported by the quantum backend. From \texttt{X} to \texttt{RZ} are \emph{standard gates} defined atomically in OpenQASM. The remaining gates from \texttt{CZ} to \texttt{C4X} are \emph{composition gates} that are constructed by standard gates. These gates are frequently used gates defined in \texttt{qelib1.inc}.

OpenQASM~2.0 is a low-level IR, which is executed sequentially without any loops, branches, or jumps, making it very convenient for static analysis and simulating in a classical simulator~\cite{li2020density, li2021sv}. A QASM code can be directly launched in IBMQ or through Qiskit. With all these benefits, QASMTrans uses QASM as the primary format for input and output.

\begin{table}[!t]
\centering\footnotesize
\caption{Basis gates for IBM-Q, Rigetti, IonQ and Quantinuum NISQ devices.}
\begin{tabular}{|c|c|c|c|}
\hline
\textbf{NISQ} & \textbf{Technology} & \textbf{1-qubit basis} & \textbf{2-qubit basis}   \\ \hline
IBMQ & Superconducting & \texttt{ID}, \texttt{RZ}, \texttt{SX}, \texttt{X} & \texttt{CX/ECR} \\ \hline
Rigetti & Superconducting & \texttt{RX}, \texttt{RZ} & \texttt{CZ} (\texttt{XY})  \\ \hline
IonQ & Trapped-Ion & \texttt{GPI}, \texttt{GPI2}, \texttt{GZ} & \texttt{MS} \\ \hline
Quantinuum & Trapped-Ion & \texttt{RX}, \texttt{RZ} & \texttt{ZZ}  \\ \hline
\end{tabular}
\label{tab:basis_gates}
\end{table}

\section{QASMTrans Transpiler} \label{sec:transpiler}
We elaborate on the QASMTrans transpiler framework in this section, The main structure is shown in Figure~\ref{fig:transpiler}. QASMTrans contains the following main components:

\begin{enumerate}
    \item \textbf{Input/Output (IO)}:
    \begin{itemize}
        \item \emph{Input}: QASMTrans starts with a QASM parser. The parser reads the QASM file, translates it into a gate IR. Meanwhile, the input module also extracts pertinent hardware details from a JSON file that describes the backend device. We plan to support other input formats such as QIR~\cite{qir} and Quil~\cite{smith2016practical}.
        \item \emph{Output}: Once the transpilation is complete, the circuit is saved to a new QASM file, primed for execution on real quantum hardware.  QIR~\cite{qir} is another format to be supported.
    \end{itemize}
    
    \item \textbf{QASMTrans Configuration}:
    \begin{itemize}
        \item \emph{Gate Decomposition}: In this phase, gates with three qubits are methodically broken down into combinations of one- and two-qubit gates. For example, the \texttt{CCX} gate will be decomposed into \texttt{CX} and \texttt{T} gates.
        \item \emph{Directed Acyclic Graph} (DAG): A DAG will be generated for the gates describing the dependency. In the DAG, every vertex represents a physical qubit, whereas each edge represents a coupling link.
        \item \emph{Coupling Graph}: We generate the coupling graph based on the input hardware JSON file, each vertex represents a physical qubit, and each edge represents the link between qubits. The coupling graph is essential for routing/mapping.
    \end{itemize}

    \item \textbf{QASMTrans Process}:
    \begin{itemize}
        \item \emph{Routing and Mapping}: This involves aligning the given quantum circuit to the specific topology of different quantum machines. To achieve this, we introduce \texttt{SWAP} gates where necessary. As a starting-point, we implement the Sabre algorithm~\cite{li+:asplos19} that is also widely used in frameworks such as Qiskit and XACC~\cite{mccaskey2020xacc}.
        \item \emph{Basis Gate Decomposition}: Depending on the desired quantum machines, like Rigetti or Quantinuum, the circuit is further decomposed into the directly executable basis gates of the specific hardware.
        
    \end{itemize}
    
    \item \textbf{Simulation-Oriented Optimization}:
    \begin{itemize}
        \item \emph{Simulation-Aware Constrained Routing}: To date, many quantum circuits and algorithms are still evaluated in classical simulators. Given the exponential cost of having more qubits to simulate, in QASMTrans, we introduce a method that can limit the number and index of qubits used for the transpilation. This can significantly reduce the transpilation time as well as simulation time.
        \item \emph{Qubit Priority Rescheduling}: Based on user-specified qubit priorities, QASMTrans can optimize and realign the qubit mapping. This is especially useful for distributive classical simulation, as the number of gates over globally shared qubits can be minimized.
    \end{itemize}
\end{enumerate}

\subsection{QASM Parser}
The QASM parser is responsible for parsing the input OpenQASM to the internal gate IR, which will be discussed in more detail below.

\subsubsection{Tokenization using Lexertk}
The parser begins its operation by tokenizing the QASM text, a process that involves breaking down the text into smaller chunks known as tokens. This is achieved by incorporating Lexertk~\cite{lexer}, a high-performance lexer tool written in C++ and distributed through a single C++ header file. The parser of QASMTrans uses Lexertk to scan through the QASM code and breaks it down into various tokens. Each token is a string of characters that conforms to the \emph{Backus–Naur Form} (BNF), an important notation technique for context-free grammars, defining a set of syntax rules for valid tokens.


\subsubsection{Qubit/Classical Register Management}
The QASM parser automatically flattens the qubit register indices, and translates them into a singular range of qubit indices. This process significantly enhances the system's proficiency for transpilation and simulation by replacing the typically used \emph{REG\_NAME[INDEX]} qubit addressing, seen in QASM, with a more streamlined one-dimensional qubit range.

Classical registers are used to store the outcomes of measurements from qubit registers, typically achieved through commands such as:

\[\text{measure }q[0] \rightarrow c[0];\]

In this example, `\emph{q}' denotes a qubit register, and `\emph{c}' denotes a classical register. The QASM parser keeps track of the qubit register remapping, ensuring accurate measurement operations.

 \begin{figure}[htb]

  \centering
  \includegraphics[width=1\linewidth]{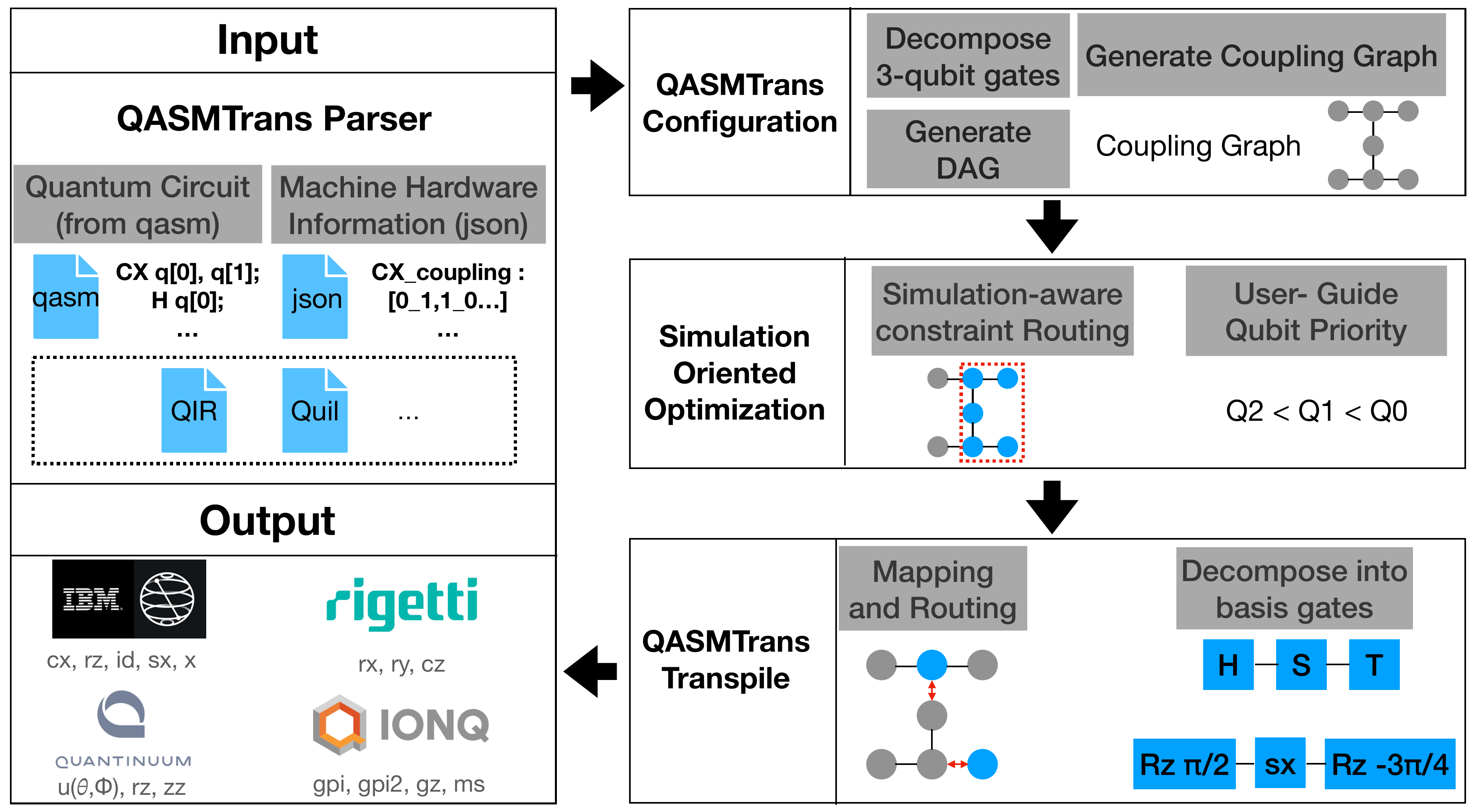}
  \caption{QASMTrans framework, which includes four major components: 1) Input/Output: the Input is the parser that reads in QASM and stores them internally as gate IRs. The Output saves the transpiled circuit in the QASM format. 
  2) Configuration: perform pre-transpilation work such as generating the coupling graph, gate DAG, and 3-qubit gates decomposition. 3) Simulation-oriented Optimization. 4) Transpilation, including mapping, routing, and decomposition into basis gates of the target device.}
  \label{fig:transpiler}
\end{figure}


\subsubsection{Gate Sets and Abstraction}
\label{sec:gate_ir} 
In the rapidly evolving field of quantum computing, it is crucial to have a robust and flexible system capable of accommodating an extensive range of quantum gates, from the most common to the more advanced. QASMTrans currently supports all the gates (except \texttt{C4X}) defined by the OpenQASM~2 specification, see Table~\ref{tab:gates}.

The parser supports standard gates such as \texttt{Pauli-X}, \texttt{Pauli-Y}, \texttt{Pauli-Z}, \texttt{Hadamard}, \texttt{CNOT}, and \texttt{Toffoli}, as well as parameterized gates like \texttt{RX}, \texttt{RY}, \texttt{RZ}, and \texttt{U} gates. It also accommodates more complex gates like the \texttt{SWAP} gate and the controlled versions of various gates. These are by no means an exhaustive list, and the parser's design allows for easy extension to incorporate additional or newer gate types.

Key to the flexibility and functionality of the QASMTrans is the Gate IR. It is a custom C++ class that encapsulates four crucial aspects of each quantum gate:

\begin{itemize}
    \item \textbf{Gate Name}: Represents the type of quantum gate.
    \item \textbf{Target Qubits}: Specifies the individual qubits upon which the quantum gate operation is performed.
    \item \textbf{Gate Parameters}: Contains the parameters relevant to certain quantum gates.
    \item \textbf{Gate Matrix}: Encapsulates the matrix representation of quantum gate, stored as two arrays --- one for the real and the other for the imaginary components.
\end{itemize}

\subsection{Transpile configuration}
Before the transpilation process, we need to perform some preliminary configuration.
\paragraph{Generation Coupling Graph (full/limited)} Based on the topology of the hardware device, we generate a coupling graph that embeds essential elements such as a distance matrix and an adjacent\_edge\_list. According to the size of the topology, there are two potential approaches: (i) Build the full graph for all the qubits and links. This, however, introduces excessive overhead towards large devices (e.g., the 433-qubit IBM Seattle). (ii) Alternatively, and in most cases, the qubit number of a  circuit is smaller than that of the device. Thus we can limit the qubits and links of the device (through a partial coupling graph) that are taken into the transpilation consideration, drastically shrinking the search space.

\paragraph{Directed Acyclic Graph (DAG) Generation} From the input circuit, a DAG can be constructed to indicate the gate dependency. For example, nodes with an in-degree of zero can be executed immediately without any dependency. Otherwise, any nodes with non-zero in-degree require all of their parent nodes to be executed beforehand to satisfy the dependency. Considering the efficiency, we only maintain two lists: one is the front list which contains executable gates; the other is the future list comprises gates for future execution.

\paragraph{Decompose three-qubit gates}  In our transpiler, we first decompose all the 3-qubit gates into 1-qubit and 2-qubit gates, given most of the quantum devices use 1-qubit and 2-qubit gates as the basis gate set. For example, the widely used \texttt{Toffoli} gate, or \texttt{CCX} gate, will be decomposed into 6 \texttt{CX} gates and 9 one-qubit gates.

\subsection{Routing and mapping}
After the initial decomposition of 3-qubit gates, the next step is to map the logical qubits to the physical qubits. Various strategies exist for performing this mapping and routing, with each method optimized for different targets. For instance, Sabre is designed to minimize the number of swaps required \cite{li+:asplos19}. Time-optimal qubit mapping emphasizes minimizing the circuit depth \cite{zhang+:asplos21}. The Noise-Adaptive approach is geared towards minimizing the error of the transpiled circuit \cite{tannu+:asplos19}. 

In QASMTrans, we use Sabre as the primary approach, due to its significant advantages in compilation time compared to the others. The major remaining overhead in Sabre routing and mapping includes:

1) After the execution of each gate, we need to update the DAG and regenerate the new front list of gates with in-degree equals to 0 in the DAG (if the gate is in the execution list, its dependency must have already been satisfied and it is ready for execution). The original Sabre method traverses the entire circuit (i.e., all DAG nodes) and identifies the gates that are ready to be executed. As QASMTrans is designed to address very deep circuits, this cost of traversing can be huge. To accelerate this process, we propose to keep the same front layer for each step, but only delete the nodes that are just executed, and fetch any new gates whose dependencies are just resolved through the step. Given that in each time step, only $n$ gates can be simultaneously executed, our proposed optimization can essentially reduce the searching cost of Sabre from $O(G)$ where $G$ is the total number of gates, to $O(n)$ where $n$ is the number of qubits. When the gate number is huge, the benefit of this improvement can be tremendous. 

2) When a \texttt{SWAP} operation is required, selecting the appropriate \texttt{SWAP} requires the calculation of all possible swaps, creating a large search space and significant overhead. This is particularly the case for large machine targets. Consequently, we propose a new method that prunes the pool of \texttt{SWAP} candidates by constraining the physical qubit area. This will be discussed in Section~\ref{subsec:simu-oriented_opt}.

\subsection{Decompose to basis gates}
Here we perform the final decomposition towards the basis gates of the device after routing and mapping. The main consideration is efficiency and simplicity, as decomposing into basis gates before routing and mapping can drastically enlarge the search space during routing and mapping. 

The decomposition here is a translation from general gates to the targeted basis gates. The basis gate set for IBMQ, Rigetti, Quantinuum and IonQ can be found in Figure~\ref{fig:transpiler}. The detailed translation rules can be found in the open-source code of QASMTrans.



\subsection{Statistics}
Based on the circuits, QASMTrans can print out the following circuit metrics based on statistics of the quantum gates in the circuit. The detailed definition can be found in \cite{li2023qasmbench}.
\begin{itemize}
\item \textbf{Circuit Depth} represents the minimum count of time-evolution steps needed to complete a quantum circuit, calculated based on standard QASM gates.

\item \textbf{Gate Density} indicates the utilization of gate slots during the time evolution of a quantum circuit, similar to pipeline occupancy in classical processors.

\item \textbf{Retention Lifespan} quantifies the maximum longevity of a qubit within a system. Its relationship with the T1 and T2 time of the device dictates the feasibility of the circuit execution on the targeted device.

\item \textbf{Measurement Density} evaluates the importance of measurement operations in a circuit, with respect to the overall induction fidelity.

\item \textbf{Entanglement Variance} measures the balance of entanglement across the qubits for a circuit. It indicates the level of connectivity and the potential error reduction through an advanced transpiler.

\end{itemize}

\subsection{Simulation-oriented Optimization} \label{subsec:simu-oriented_opt}

As mentioned, most of the contemporary circuit inductions are still performed through classical simulations. In QASMTrans, we propose two classical simulation-oriented optimizations during transpilation to generate circuits that can be simulated more efficiently. 

\paragraph{Constrained qubit routing/mapping} During the routing and mapping phase, instead of considering all the physical qubits of the device, we limit the number and coupling of qubits that will be considered during the transpilation, based on the number of logical qubits used in the circuit. This is achieved by first adopting the isomorphic algorithm to find the most relevant connected graph from the hardware architecture, using the number of logical qubits as input. The qubits of the obtained graph should contain equal or more qubits than the circuit logical qubits, but less or equal to the number of physical qubits in the device. We then refer to the routing algorithm as normal. Although constrained routing and mapping with partial graphs can lead to more swaps, the benefit of simulating fewer qubits can extraordinarily speed up the transpilation process.

\paragraph{User-guided qubit prioritization}
Another simulation-oriented optimization is to enforce user-defined qubit prioritization. Users can specify a priority order such as $q3<q1<q0<q2$, then for classical simulation, we can perform a qubit remapping with respect to this partial order. This is achieved by counting the number of gates performed on each qubit, sorting, and then re-indexing the qubits to assign high-priority qubits to perform more gates. For example, if $q2$ shows the best performance or least error rate, which is set to have the highest priority, the qubit with the most number of gates can be remapped to it. On the other hand, if the coefficients of $q3$ are distributed across multiple nodes for large-scale distributive simulation (i.e., a global qubit), because of the overwhelming cost from inter-node communication, it is set to the lowest priority, we would want the least number of gates to be mapped to $q3$.

\section{Evaluation} \label{sec:evaluation}

\begin{table*}
\centering
\refstepcounter{table}
\caption{Evaluation of QASMTrans compared to Qiskit and Qmap in terms of transpilation quality and efficiency. "X" implies no results obtained with an hour of transpilation time.}
\label{tab:Compilation_time_analysis}
\begin{tabular}{|c|c|c|c|c|l|l|l|c|c|c|} 
\hline
\multirow{2}{*}{Benchmarks} & \multicolumn{4}{c|}{\textbf{Input} Circuit Information} & \multicolumn{3}{c|}{\textbf{Quality}: Transpiled by Qiskit/Qmap/QASMTrans~} & \multicolumn{3}{c|}{~\textbf{Efficiency}: Transpilation Time~}  \\ 
\cline{2-11}
                            & Qubits & 2q Gate & Total Gates & Depth  & 2q Basis Gates & Total Basis Gates    & Transpiled Depth~            & Qiskit  & Qmap    & QASMTrans              \\ 
\hline
square\_root           & 18     & 898     & 2300  & 1268   & 2663/2742/2911 & 4079/4300/5665 & 2663/2892/3112    & 1210 ms & 392 ms  & 39 ms                  \\ 
\hline
vqe\_uccsd               & 8      & 5488    & 10808 & 7252   & 5975/5975/6070 & 12K/19K/26K    & 9K/10K/11K        & 4620 ms & 2125 ms & 154 ms                 \\ 
\hline
vqe\_uccsd              & 24     & 1.5M    & 2.2M  & 1M     & X/X/2.7M       & X/X/7.7M       & X/X/2.8M          & X       & X       & 69 s                   \\ 
\hline
sat                    & 11     & 252     & 679   & 409    & 739/710/739    & 1118/877/1229  & 879/712/879       & 715 ms  & 102 ms  & 8 ms                   \\ 
\hline
bwt                    & 21     & 174K    & 470K  & 222K   & 648K/619K/640K & 885K/812K/998K & 598K/564K/601K    & 2544 s  & 299 s   & 6.9 s                  \\ 
\hline
gcm                     & 13     & 762     & 3148  & 2447   & 1376/1168/1422 & 3706/3554/3808 & 2955/2654/2962    & 5.2 s   & 652 ms  & 40 ms                  \\ 
\hline
hhl                    & 10     & 72K     & 186K  & 147K   & 95K/93K/103K   & 241K/236K/312K & 213K/211K/258K    & 782 s   & 291 s   & 2.4 s                  \\ 
\hline
qaoa                    & 6      & 54      & 270   & 109    & 96/96/96       & 887/906/1032   & 205/257/243       & 648 ms  & 148 ms  & 4.2 ms                 \\ 
\hline
qec                         & 5      & 10      & 25    & 22     & 22/22/22       & 48/37/49       & 25/22/25          & 193 ms  & 61 ms  & 5 ms                   \\ 
\hline
adder                       & 4      & 10      & 23    & 11     & 16/16/16       & 33/33/33       & 18/18/18          & 203 ms  & 98 ms   & 4.9 ms                 \\ 
\hline
adder                       & 10     & 65      & 142   & 99     & 146/110/146    & 278/219/287    & 243/205/249       & 396 ms  & 155 ms  & 6.6 ms                 \\ 
\hline
adder                       & 64     & 455     & 988   & 369    & 2660/2198/2660 & 3494/3255/3641 & 2550/1958/2584    & 13.2 s  & 6.7 s   & 932 ms                 \\ 
\hline
bv                          & 140    & 72      & 352   & 75     & 444/434/496    & 1281/1192/1468 & 307/298/412       & 8.9 s   & 2.5 s   & 496 ms                 \\ 
\hline
ghz                         & 255    & 254     & 255   & 255    & 797/797/813    & 802/768/818    & 802/798/818       & 7.9 s   & 2.1 s   & 955 ms                 \\ 
\hline
qft                         & 320    & 102K    & 255K  & 2549   & X/X/767K       & X/X/1.5M       & X/X/806K          & X       & X       & 31 s                   \\ 
\hline
ising                       & 420    & 838     & 3614  & 16     & 1382/1320/1672 & 5062/5012/5089 & 36/36/36          & 1910 ms & 491 ms  & 59 ms                  \\
\hline
\end{tabular}
\end{table*}

\subsection{Experimental setup}
We mainly use the NERSC Perlmutter HPC system for the evaluation. Perlmutter is built by HPE. Each of the Cray EX systems is equipped with an AMD EPYC 7763 CPU and four NVIDIA A100 GPUs. The other platforms used for the transpilation are listed in Table~\ref{tab:platform}. We compare QASMTrans to two state-of-the-art and most relevant quantum transpilers for comparison: Qiskit~\cite{IBMQiskit} (with Sabre algorithm~\cite{li+:asplos19}) and MQT-Qmap~\cite{wille2023mqt}. We focus on transpilation efficiency, quality, and fidelity. The efficiency is measured by transpilation time. The quality is measured by the depth, total number of gates, and number of \texttt{CX} gates of the transpiled circuit. The fidelity is measured by calculating the fidelity of execution for the transpiled circuit over five real quantum devices: \emph{IBM-Brisbane}, \emph{Rigetti-AspenM2}, \emph{IonQ-Aria1} and \emph{Quantinuum-H1-1}). We test on different benchmark circuits varying from 10 qubits to 400 qubits from QASMBench~\cite{li2023qasmbench}.

\subsection{Transpilation Efficiency and Quality}

The evaluation results are listed in Table~\ref{tab:Compilation_time_analysis}. We use IBMQ devices as the transpilation target so that: (i) the basis gate set is \texttt{X}, \texttt{SX}, \texttt{CX}, and \texttt{RZ}; (ii) for topology, when the number of qubits of the circuit is less than 27, we use the topology of IBMQ Toronto. When it is larger than 27, we use the topology of the latest 433-qubit IBM Seattle as the objective device.

\vspace{4pt}\noindent\textbf{Quality:} Overall, QASMTrans can generate transpiled circuits with comparable depth, gates and 2-qubit gates as Qiskit and Qmap. The slight difference is due to the fact that as the initial effort, QASMTrans hasn't yet implemented or integrated advanced front-end gate transformation \& cancellation passes.  

\vspace{4pt}\noindent\textbf{Efficiency:} As listed, QASMTrans shows a tremendous performance advantage over Qiskit and Qmap for the 16 benchmark circuits. The speedup can be as much as 369$\times$ over Qiskit and 61$\times$ over Qmap. In particular, for some challenging circuits, such as the vqe\_uccsd\_n24 with 2.2M gates, and qft\_n320 with 255K gates, neither Qiskit nor Qmap can produce transpiled circuits within a reasonable time (i.e., 1 hour), while QASM can accomplish in 69s and 31s, respectively.

\begin{figure}[htb]

  \centering
  \includegraphics[width=0.95\linewidth]{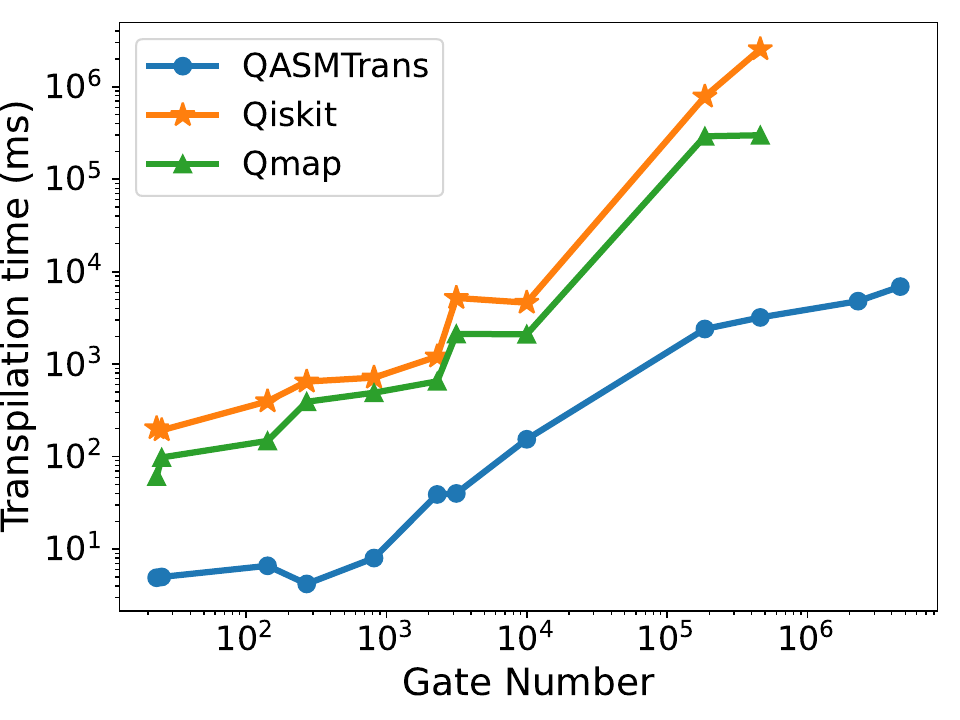}
  \caption{Transpilation time with respect to the number of gates of the input circuits. The advantage of QASMTrans over Qiskit and Qmap is consistent. The last two points are for \emph{vqe\_uccsd\_n24} and \emph{qft} where Qiskit and Qmap cannot finish transpilation in 1 hour, see Table~\ref{tab:Compilation_time_analysis}.}
  \label{fig:gate_number}
\end{figure}

\vspace{4pt}\noindent\textbf{Scalability:} We further look at the performance scalability. Figure~\ref{fig:gate_number} shows the scaling of the transpilation time with respect to the number of gates of the input circuits for the various benchmarks. As can be seen, the performance advantage over Qiskit and Qmap is quite consistent.

\subsection{Transpilation Fidelity}
\begin{figure}[htb]
  \centering
  \includegraphics[width=1\linewidth]{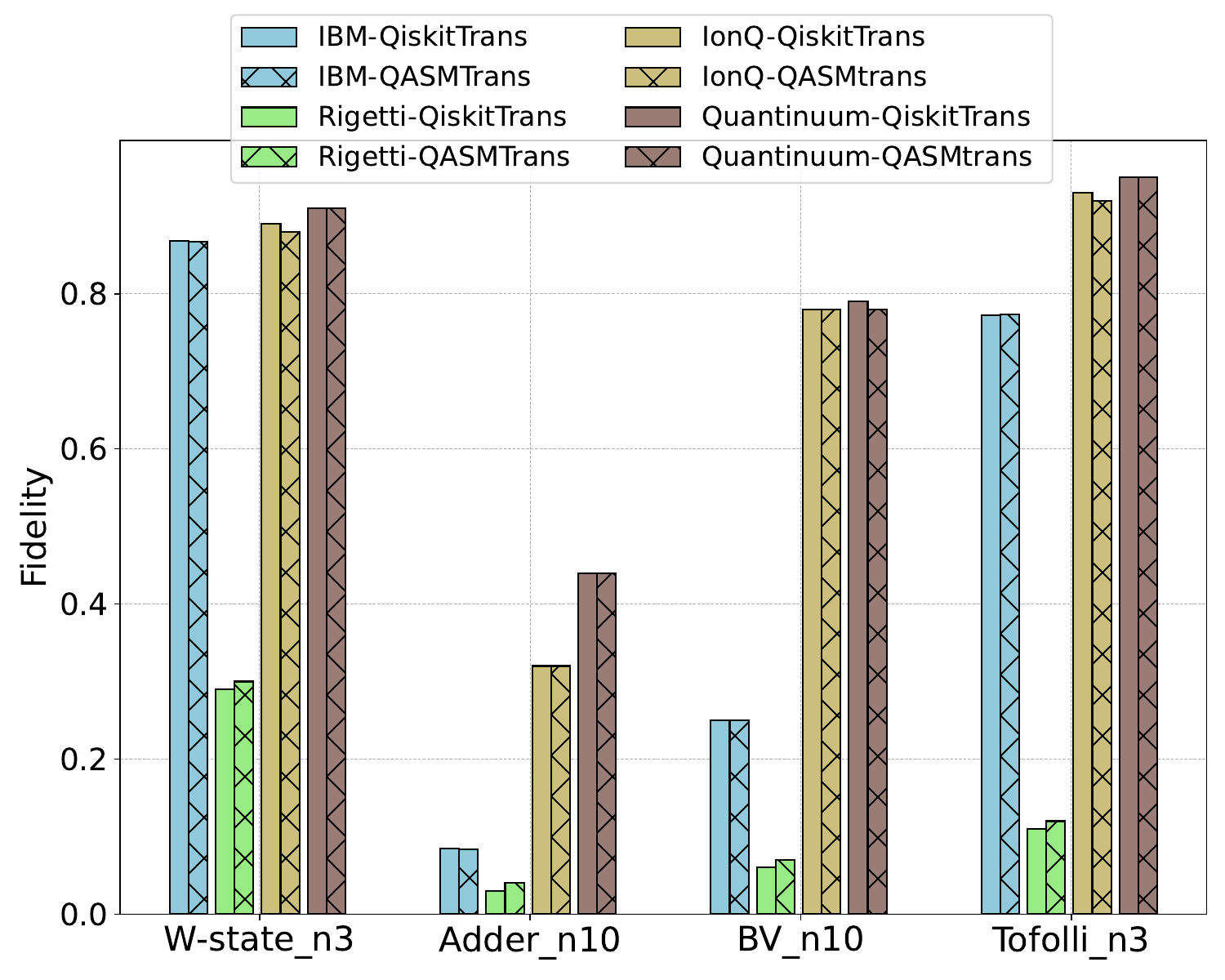}
  \caption{QASMTrans fidelity analysis compared with Qiskit transpiler on different machines, the X-axis shows the benchmarks, and Y-axis shows the fidelity obtained on real NISQ machines (IBM \emph{ibm\_brisbane}, Rigetti \emph{Aspen-M2}, IonQ \emph{Aria-1}, and Quantinuum \emph{H2}) with respect to the Qiskit results.}
  \label{fig:fidelity}
\end{figure}

To evaluate the correctness of transpilation, we use the transpiled circuits generated by Qiskit and QASMTrans as the inputs, and launch them onto four real NISQ devices (IBMQ, Rigetti, Quantinuum, and IonQ) to assess the difference in their induction results, shown in Figure~\ref{fig:fidelity}. Please be aware that these input circuits, despite having already been transpiled, may go through another round of internal transpilation or optimization within the backend processing of the NISQ device. This is not under our control. However, we argue that this will not significantly impact the fidelity results since both input circuits go through the same backend processes.

As can be seen in Figure~\ref{fig:fidelity}, the fidelity with Qiskit result is quite consistent across input circuits and underlying hardware, with $<1\%$ deviation. This underscores the robustness and stability of QASMTrans.

\subsection{Optimization for Classical Simulation}

Both the constrained qubit routing/mapping and user-guided qubit prioritization presented in Section~III-F can harvest performance gain for classical simulation. Constrained qubit routing/mapping limits the number of qubits for the simulation, for which the performance gain is quite obvious. Here, we mainly focus on demonstrating the benefit of user-guided qubit prioritization.

\begin{figure}[htb]
  \centering
  \includegraphics[width=0.95\linewidth]{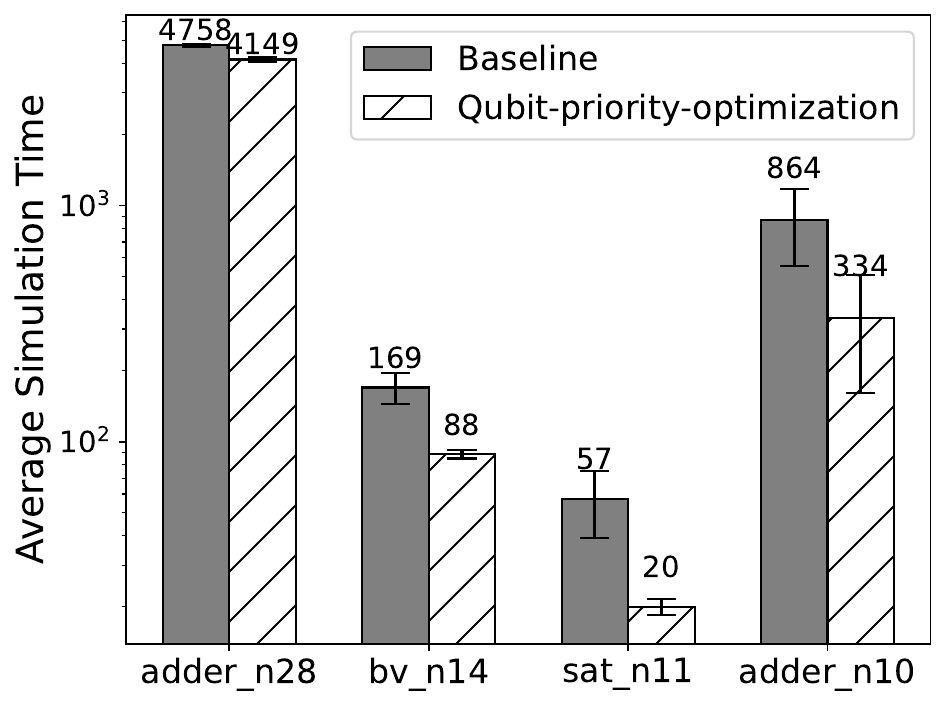}
  \caption{Performance gain in classical simulation through user-guided qubit prioritization using SV-Sim for the transpiled circuits on 8 GPUs of Perlmutter. Note, the Y-axis of simulation time is in log-scale.}
  \label{fig:remap}
\end{figure}

We have already discussed why minimizing the number of gates over the global qubits can reduce the overhead from communication. Here, we use SV-Sim~\cite{li2021sv} as the classical simulator. We use all the 8 GPUs from 2 Perlmutter nodes for the distributive circuit simulation. Consequently, 3 qubits are sharing their corresponding coefficients across the 8 GPUs. Figure~\ref{fig:remap} shows the difference in simulation time for the transpiled circuits with and without user-guided qubit prioritization. As can be seen, the performance gain can be quite significant given the log-scale of the Y-axis. This benefit mainly comes from switching some expensive gates over the three global qubits to local qubits through the final remapping of qubit prioritization.

\subsection{Platform Portability}
\begin{figure*}[htb]

  \centering
  \includegraphics[width=0.95\linewidth]{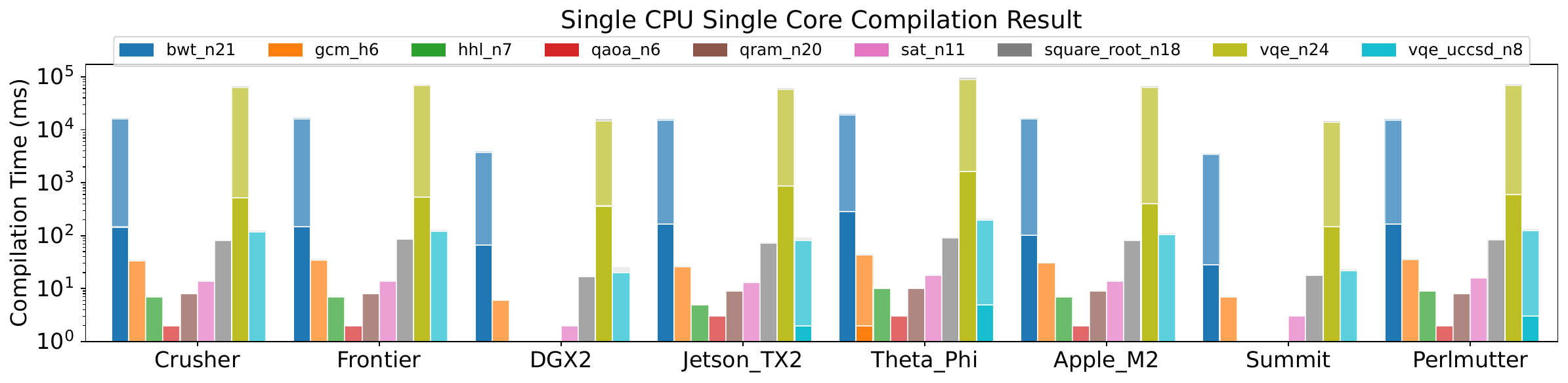}
  \caption{Compilation time on various platforms. The X-axis shows the name of different platforms, while the Y-axis is the compilation time using a single core of a CPU of the system. The empty bars indicate the condition that the compilation time is less than 1ms. The breakdown of each bar implies the time of (upper) routing \& mapping, and (lower) decomposition. Please be aware that the Y-axis is in the log scale.
 }
  \label{fig:fidelity_platform}
\end{figure*}

We evaluate QASMTrans across different computing platforms, from various HPC systems, including NERSC Perlmutter, OLCF Frontier, Crusher, and Summit, ALCF Theta, to a desktop and laptop (Intel P8168 and Apple M2), to an embedded device (JetsonTX2 with ARM8). The platforms are listed in Table~\ref{tab:platform}. The results are shown in Figure~\ref{fig:fidelity_platform}. The transpilation on all the platforms can be finished within 100s and most of them below 1s.

\begin{table}
\caption{Platforms for Portability Evaluation}
\centering
\begin{tabular}{|c|c|c|c|c|c|} 
\hline
Platform                                               & CPU                                                            & Vendor & Core & Mem   & Compiler                                                     \\ 
\hline
\begin{tabular}[c]{@{}c@{}}MacBook \\Pro\end{tabular} & Apple M2                                                             & Apple  & 12   & 16GB  & \begin{tabular}[c]{@{}c@{}}AppleClang\\~14.0.3\end{tabular}  \\ 
\hline
Perlmutter                                             & \begin{tabular}[c]{@{}c@{}}Authentic\\AMD\end{tabular}         & AMD    & 128  & 256GB & g++ 11.2.0                                                   \\ 
\hline
JetsonTX2                                              & ARMV8                                                          & NVIDIA & 4    & 8GB   & g++ 5.4.0                                                    \\ 
\hline
Crusher                                                & \begin{tabular}[c]{@{}c@{}}Authentic\\AMD\end{tabular}         & AMD    & 128  & 512GB & g++ 12.2.0                                                   \\ 
\hline
Frontier                                               & \begin{tabular}[c]{@{}c@{}}Authentic\\AMD\end{tabular}         & AMD    & 128  & 512GB & g++ 12.2.0                                                   \\ 
\hline
Summit                                                 & POWER9                                                         & IBM    & 176  & 512GB & g++ 9.1.0                                                    \\ 
\hline
Tonga                                                  & Intel P8168                                                    & Intel  & 96   & 128GB & g++ 11.2.0                                                   \\ 
\hline
Theta                                                  & \begin{tabular}[c]{@{}c@{}}Intel Phi \\7230 (KNL)\end{tabular} & Intel  & 256  & 192GB & intel 19.1.0                                                 \\
\hline
\end{tabular}
\label{tab:platform}
\end{table}

With these results, we have three observations: (i) QASMTrans can be portable on various platforms, given its efficient C++ based implementation and non-external library dependency (the \emph{json} and \emph{lexertk} are included as header files). In particular, the successful and efficient running on an ARM8 CPU shows the potential of practical deployment on an FPGA of a real quantum system or testbed, such as LBNL AQT. (ii) The transpilation speed across applications circuits and platforms is consistent. (iii) The majority (nearly $90\%$) of the transpilation time are devoted to routing and mapping for the current implementation of QASMTrans.

\section{Related Work} \label{sec:related_work}

\subsection{Quantum Intermediate Representation}

In quantum computing, gate IR provides an essential abstraction layer, offering a structured, machine-agnostic representation of quantum circuits. Among the existing quantum IRs, the Microsoft QIR~\cite{qir} is an LLVM-based IR that defines a set of rules for representing quantum constructs. QIR attempts to serve as a common interface between various quantum languages (e.g., Q\#) and platforms. QASM~\cite{cross2017open} is a widely recognized quantum assembly language developed by IBM for its hardware platforms and software tool-chain. Quil~\cite{quil} is a portable quantum instruction language developed by Rigetti. Lastly, XACC (eXtreme-scale ACCelerator)~\cite{mccaskey2020xacc} is a compilation framework for hybrid quantum-classical computing architectures developed at ORNL, supporting IBM, Rigetti, D-Wave QPUs and various classical simulators such as SV-Sim~\cite{li2021sv} and DM-Sim~\cite{li2020density}.


\subsection{Quantum Transpilation}

Quantum transpiler plays a crucial role in quantum computing by translating high-level quantum algorithms into a series of low-level hardware-specific instructions that quantum hardware can execute.
Qiskit is a widely used quantum software development package developed by IBM. The Qiskit transpiler provides a flexible and extensible framework, offering a wide array of compilation passes that can be combined in different ways to create customized and hardware-tailored transpilation pipelines. 

In addition to Qiskit, there are various transpilers aiming at different purposes:
1) \emph{application-oriented transpilation:} These transpilers focus on specific domain applications. For example, Paulihedral~\cite{li+:asplos22} focuses on VQE, Twoqan~\cite{lao+:arxiv21twoqan} concentrates on QAOA circuits.
2) \emph{hardware-oriented transpilation:} These transpilers focus on supporting the new features of a particular quantum platform. For instance, CaQR emphasizes the support for dynamic circuit generation and the opportunities from qubit reset~\cite{hua2023caqr}. Pulse transpilers delve into the nuances of low-level pulse scheduling, optimizing quantum operations at the physical layer \cite{gokhale2020optimized, chen2023pulse, shi+:asplos19}. AutoComm~\cite{wu2022autocomm} and QuComm~\cite{wu2022collcomm} present transpiler optimization techniques for distributive quantum devices.
3) \emph{Optimization for mapping/routing:} there are a bunch of works aiming at improving general transpilation performance, like Sabre~\cite{li+:asplos19} and Zulhner ~\cite{zulehner+:date18:} attempt to minimize the number of additional gates in mapping/routing. TOQM \cite{zhang+:asplos21} aims at shrinking the depth of the transpiled circuit. Shi {\it et al.}~\cite{shi+:asplos19} presents the complete transpilation and optimization flow, including gate aggregation and cancellation. QASMTrans falls into the third category, aiming at improving the transpilation performance of large and deep QASM circuits.



\section{Conclusion} \label{sec:conclude}

In this paper, we present QASMTrans, a C++ based quantum transpiler framework for NISQ devices. It outperforms prevalent counterparts, notably achieving up to more than 300$\times$ speedups over the Qiskit transpiler. We demonstrate the quality, efficiency, and fidelity of QASMTrans across various classical and quantum platforms. Future work includes continuously improving QASMtrans by adding new passes such as gate cancellation, new platform support such as for distributed quantum computing and cavity-based systems, as well as the support of new input/output formats such as QIR.

\section*{Acknowledgement}

This material is mainly based upon work supported by the U.S. Department of Energy, Office of Science, National Quantum Information Science Research Centers, Co-design Center for Quantum Advantage (C2QA) under contract number DE-SC0012704. The contribution from Meng Wang, Yufei Ding, and Travis Humble are supported by the U.S. Department of Energy, Office of Science, National Quantum Information Science Research Centers, Quantum Science Center (QSC). This research used resources of the Oak Ridge Leadership Computing Facility, which is a DOE Office of Science User Facility supported under Contract DE-AC05-00OR22725. This research used resources of the National Energy Research Scientific Computing Center (NERSC), a U.S. Department of Energy Office of Science User Facility located at Lawrence Berkeley National Laboratory, operated under Contract No. DE-AC02-05CH11231. We acknowledge support from Microsoft's Azure Quantum for providing credits and access to the ion-trap quantum hardware.
The Pacific Northwest National Laboratory is operated by Battelle for the U.S. Department of Energy under Contract DE-AC05-76RL01830.

\bibliographystyle{IEEEtranS}
\bibliography{refs,references}

\end{document}